\documentclass[prd,superscriptaddress,amsfonts,amssymb,amsmath,showpacs,twocolumn]{revtex4-2}
\usepackage{bm}
\usepackage{amsfonts}
\usepackage{latexsym}
\usepackage{graphicx}
\usepackage{amsmath}
\usepackage{palatino}
\usepackage{mathpazo}
\usepackage{textcomp}
\usepackage{float}
\usepackage{booktabs}
\usepackage{dcolumn}
\usepackage{booktabs}
\usepackage{multirow}
\usepackage{hyperref}
\hypersetup{colorlinks,citecolor=blue}
\usepackage{amsmath}
\usepackage{xcolor}
\usepackage{orcidlink}
\usepackage{epsfig}
\usepackage{caption}
\usepackage{subcaption}
\usepackage{commath}
\hypersetup{colorlinks,citecolor=blue}
\hypersetup{colorlinks=true,linkcolor=magenta,filecolor=magenta,    urlcolor=blue}

\def\be{\begin{equation}}
\def\ee{\end{equation}}
\def\bea{\begin{eqnarray}}
\def\eea{\end{eqnarray}}

\begin{document}
\title{ Joule-Thomson expansion and tidal force effects of AdS black holes surrounded by Chaplygin dark fluid}

\author{Dhruv Arora}
\email{arora09dhruv@gmail.com}
\affiliation{Pacif Institute of Cosmology and Selfology (PICS), Sagara, Sambalpur 768224, Odisha, India}
\author{Muhammad Yasir}
\email{yasirciitsahiwal@gmail.com}\affiliation{Department of Mathematics, Shanghai University,
Shanghai, 200444, Shanghai, People's Republic of China}
\author{Himanshu Chaudhary}
\email{himanshuch1729@gmail.com} 
\affiliation{Department of Applied Mathematics, Delhi Technological University, Delhi-110042, India}
\affiliation{Pacif Institute of Cosmology and Selfology (PICS), Sagara, Sambalpur 768224, Odisha, India}
\affiliation{Department of Mathematics, Shyamlal College, University of Delhi, Delhi-110032, India.}
\author{Faisal Javed}
\email{faisaljaved.math@gmail.com}\affiliation{Department of Physics, Zhejiang Normal University, Jinhua 321004, People's Republic of China}
\author{G. Mustafa}
\email{gmustafa3828@gmail.com}
\affiliation{Department of Physics,
Zhejiang Normal University, Jinhua 321004, People’s Republic of China}
\affiliation{Zhejiang Institute of
Photoelectronics and Zhejiang Institute for Advanced Light Source,
Zhejiang Normal University, Jinhua, Zhejiang 321004, China}
\author{Xia Tiecheng}
\email{xiatc@shu.edu.cn}
\affiliation{Department of Mathematics, Shanghai University  and Newtouch Center for Mathematics of Shanghai University,  Shanghai ,200444, P.R.China}
\author{Farruh Atamurotov}
\email{atamurotov@yahoo.com} 
\affiliation{New Uzbekistan University, Movarounnahr street 1, Tashkent 100000, Uzbekistan}
\affiliation{Central Asian University, Milliy Bog' Street 264, Tashkent 111221, Uzbekistan}
\affiliation{University of Tashkent for Applied Sciences, Str. Gavhar 1, Tashkent 100149, Uzbekistan}
\affiliation{Institute of Theoretical Physics, National University of Uzbekistan, Tashkent 100174, Uzbekistan}

\begin{abstract}
This study examines a recently hypothesized black hole.  We study the Joule-Thomson coefficient, the inversion temperature and also the isenthalpic curves in the $T_i -P_i$ plane. A comparison is made between the Van der Waals fluid and the black hole to study their similarities and differences.  The Joule-Thomson coefficient, the inversion curves and the isenthalpic curves are discussed inAdS black holes surrounded by Chaplygin dark fluid.  In $T -P$ plane, the inversion temperature curves and isenthalpic curves are obtained with different parameters. Next, we explore the radial timelike geodesics that leads us to explore the tidal force effects for a radially in-falling particle in such black hole spacetime. We also numerically solve the geodesic deviation equation for two nearby radial geodesics for a freely falling particle. Our analysis shows that contrary to the Schwarzschild spacetime, the tidal forces don't become zero at spatial infinity due to the lack of asymptotic flatness because of the presence of a non-zero cosmological constant. The geodesic separation profile shows an oscillating trend and depends on the dynamic spacetime parameters $q, B$ and $\Lambda$.
\end{abstract}
\maketitle
\section{Introduction}
The theory of general relativity has also played a crucial role in explaining the behavior of massive objects, such as black holes (BHs) and the expansion of the universe. Its predictions have been confirmed by numerous experiments and observations, solidifying its status as a cornerstone of modern physics. The physical interpretations of Einstein's gravitational field equation solutions are crucial to the development of general relativity. Therefore, the solutions to these equations and the examination of their properties are essential components of the theory of gravity. Every gravitational solution contributes significantly to our understanding of spacetime and gravity. Furthermore, the investigation of high-dimensional gravitational solutions advances our knowledge of the dimensional dependence and stability of gravitational characteristics in spacetime. The exact solution of Einstein field equations referred as BH is a direct outcome of the theory of general relativity. The event horizon of BH  behaves as a barrier that distinguishes between the interior and exterior regions of a BH. Once particles traverse this barrier, their escape becomes impossible. In the last several decades, a significant body of research \cite{P1,P2,P3} has demonstrated a fundamental and crucial correlation between BHs and general thermodynamic systems. Additionally, BHs may be seen as thermodynamic systems with measurable temperature and entropy. Consequently, the existence of BHs has a profound influence on our comprehension of quantum gravity, as evidenced by scholarly works \cite{Q1,Q2,Q3,Q4}.\\\\
Enormous scholarly attention has been dedicated to investigating the thermodynamic characteristics of BHs following a substantial period of scientific advancement. This encompasses inquiries into the impact of quantum phenomena on the thermodynamics of BHs \cite{ch1,ch2,ch3,ch4}, the concept of entropy \cite{EE1,EE2}, transitions between different thermodynamic phases \cite{PT1,PT2,PT3}, the emission of both thermal and non-thermal radiations \cite{R1,R2,R3}, and other critical phenomena \cite{C1,C2,C3,C4,C5,C6,C7}. In the context of the extended phase space, the negative cosmological constant is seen as a variable thermodynamic pressure within a system of BHs. Its corresponding conjugate quantity is readily interpreted as the volume of the BH in the mechanics of Anti-de Sitter (AdS) BHs. The mass of a BH is commonly understood to represent the enthalpy of spacetime rather than its intrinsic energy \cite{A1}. The aforementioned notions have been widely extended and used in the context of BHs within the framework of the Lovelock theory \cite{A2}.
Moreover, a scholarly investigation of dilaton-AdS black holes has demonstrated that the mass of a BH is indicative of its enthalpy  \cite{A2},\cite{A4}. The aforementioned investigations have yielded significant outcomes in the field of thermodynamics study pertaining to AdS BHs inside the extended phase space.\\\\
It is possible to deduce thermodynamic aspects of certain classical systems by the utilization of similarities between the behavior of BHs \cite{M1,M2} and the characteristics exhibited by van der Waals fluids \cite{F1,F2,F3}. The presence of commonalities across BHs in AdS space implies a substantial correlation between AdS and CFT \cite{ADS}.  Furthermore, there has been much expansion of other ideas, such as those pertaining to revolving and hairy BHs \cite{BH1,BH2}, as well as m-theory \cite{MT1,MT2}. In order to accentuate the similarities with van der Waals fluids, researchers have achieved further noteworthy findings, including the examination of quasi-normal modes \cite{QU1,QU2}, the investigation of holographic heat engines \cite{HE1}, and the exploration of chaotic structures \cite{ST1}. Recently, the study of thermal analysis with quantum corrections and emission energy of charged BH with nonlinear electrodynamics are explored in \cite{fa1}. The study various BHs thermodynamics and phase transitions are discussion in \cite{fa3,fa4,fa5,fa6}.\\\\
The phenomenon of gas flow from a region of higher pressure to a region of lower pressure at an equivalent rate, as seen in classical thermodynamics, is often known as Joule-Thomson (J-T) expansion. In previous research, \"Okc\"u and E. Aydiner  \cite{J1} conducted an investigation on the charged AdS BHs, specifically focusing on the J-T effect.  Numerous comprehensive research have provided empirical evidence supporting the precise identification of heating and cooling areas through the utilization of the temperature-pressure ($T$-$P$) diagram. The point at which the inversion and enthalpic curves overlap provides a means of distinguishing between the zones of heating and cooling within the BH. When the J-T coefficient $\mu$ is positive, it corresponds to the cooling zone, whereas a negative value of $\mu$ indicates the heating region. Furthermore, the relationship between the minimum temperature ($T_{min}$) and the critical temperature ($T_c$) is of importance in the Joule-Thomson expansion. Specifically, for a Reissner-Nordström (RN) AdS, this ratio is determined to be $1/2$ according to reference \cite{J1}. Prior research has examined this proportion for additional prominent BHs \cite{J2,J3,J5}. This study has restored the ratio to one-half in specified cases. Following this, the research of J-T growth gained prominence and received considerable attention, encompassing the examination of several categories of BHs \cite{J2,J3,J4,J5,J6,J7,fa2, J8}.\\\\ 
It is widely acknowledged that within the framework of Schwarzschild spacetime, the motion of a falling object towards the event horizon is subject to radial tidal forces that cause stretching, while compression occurs in the angular directions \cite{e1,e2}. Recent research has focused on analyzing tidal forces in different types of BHs, including RN BH \cite{e3}, Kiselev \cite{e4}, and other regular BHs \cite{e5}. Tidal forces in Kerr spacetime have also been investigated, as evidenced by references \cite{e6}-\cite{e11}. These forces play a critical role in astrophysics, as they can result in the disruption of a star due to the tidal forces exerted by a BH \cite{e6}-\cite{e11}, leading to the occurrence of a tidal disruption event (TDE). TDEs can produce luminous flares in the form of x-ray \cite{e12}, ultraviolet \cite{e13}, and optical \cite{e14}  radiation.\\\\
In this paper, we are interested to generalize the current research on the Joule-Thomson expansion for AdS BHs surrounded by Chaplygin dark fluid and brief study on tidal forces. The formation of the current paper is as written. In Sec. \ref{sec2}, we study a brief review of AdS BHs surrounded by Chaplygin dark fluid. Then, we explore the J-T expansion in Sec. \ref{sec3}. Sec. \ref{sec4} and \ref{sec5} is related to the radial geodesics and tidal forces, respectively. Sec. \ref{sec6} and \ref{sec7} present the radial as well as angular tidal force and geodesic deviation, respectively.  Finally, we present a few closing remarks in the last section.
\section{AdS black hole surrounded by dark fluid with Chaplygin-like equation of state}~\label{sec2} 
\begin{figure}[htbp]
\centering
\includegraphics[scale=0.51]{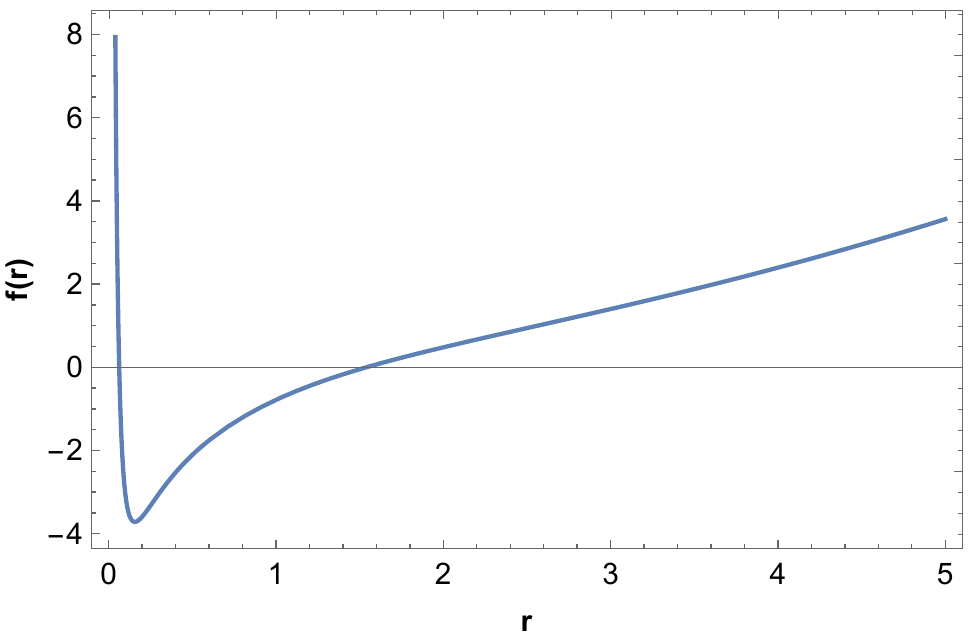}
\caption{Plot of the lapse function w.r.t the radial coordinate $r$ depicting the location of horizons for $q = 0.743$, $B = 0.772$ and $\Lambda = -1.234$}
\label{fig:1}
\end{figure}

Exploring the interaction between a BH and the surrounding field involves solving gravitational field equations alongside the equation of motion for the relevant field. This is particularly practical when dealing with a static spherically symmetric BH and its atmosphere composed of a field with a specific Lagrangian. Understanding the nature of the matter near BH solutions becomes challenging, especially in the presence of quintessence dark energy and Chaplygin gas. In such cases, it becomes essential to explore the relationship between matter and spacetime curvature solely based on the equation of state for the fluid matter. In our current investigation, we adopt a specific metric form to characterize the static spherically symmetric spacetime.
\begin{equation}\label{eq1}
ds^{2}=-f(r)dt^{2}+\frac{1}{f(r)}dr^{2}+r^{2}(d\theta^{2}+ \sin^{2}\theta d\phi^{2}).
\end{equation}
To describe the matter source, we initiate with a perfect fluid, which is given as
\begin{equation}\label{eq2}
T_{ab}=pg_{ab}+ (\rho+p) u_a u_b, 
\end{equation}
In the field Semiz~\cite{jim1} made significant contributions by investigating the equation of state (EOS) \(p = \omega \rho\) (\(\omega\) being a constant) for ideal fluid sources like dust, phantom energy, radiation, or dark energy, employing Einstein's equations as the foundational framework. Another notable contribution came from Kiselev~\cite{jim2}, who derived a novel BH spacetime by considering ambient quintessence matter as an anisotropic fluid, a model that gained widespread popularity. The understanding of the background processes leading to the Chaplygin dark fluid (CDF) remains uncertain from a field theoretical standpoint~\cite{Hulke2020,Ray2023}, although some potential explanations exist within the framework of string theory~\cite{Nozari2011,Benaoum2012}. The CDF is often described by introducing a self-interacting potential and a scalar field \(\varphi\) with the Lagrangian \(L_{\varphi} = -\frac{1}{2}\partial_{a}\varphi\partial^{a}\varphi-U(\varphi)\)~\cite{jim3,jim4}. This study assumes that the CDF possesses anisotropic properties, and its stress-energy tensor can be expressed covariantly~\cite{jim5}.
\begin{equation}\label{eq3}
T_{ab}= \rho u_a u_b +p_r k_a k_b +p_t \Pi_{ab}, 
\end{equation}
The radial and tangential pressure components are represented by \(p_r\) and \(p_t\), respectively. The four-velocity of the fluid is denoted by \(u_a\), and \(k_a\) represents a unit space-like vector that is orthogonal to \(u_a\). It is worth noting that \(u_a\) and \(k_a\) satisfy the conditions \(u_a u^a = -1\), \(k_a k^a = 1\), and \(u^a k_a = 0\). The tensor \(\Pi_{ab} = g_{ab} + u_a u_b - k_a k_b\) serves as a projection onto the two orthogonal surfaces defined by \(u^a\) and \(k^a\). When working in the fluid's co-moving frame, specific values can be assigned, such as \(u_a = (-\sqrt{f},0,0,0)\) and \(k_a = (0,1/\sqrt{g},0,0)\). Finally, the stress-energy tensor, as derived from Eq.~(\ref{eq3}), is expressed as follows.
\begin{equation}
T_{a}{}^{b}=-(\rho+p_t)\delta_{a}{}^0\delta^{b}{}_0 +p_t \delta_{a}{}^{b} +(p_r-p_t)\delta_{a}{}^1\delta^{b}{}_1. \label{eq4}
\end{equation}
The anisotropic factor is expressed as the disparity between radial and tangential pressures, denoted as \(p_r - p_t\). In situations where \(p_r = p_t\), the stress-energy tensor simplifies to the standard isotropic form. Even if a cosmological fluid exhibits anisotropy in the gravitational field produced by a black hole, its equation of state should manifest as \(p = p(\rho)\) on a cosmological scale. This allows for the imposition of constraints on the tangential pressure (\(p_t\)) by performing an isotropic averaging over angles and incorporating additional conditions. $\langle {T}_i{}^j\rangle=p(\rho)\delta_i{}^j$, An individual can obtain the following relationship
\begin{equation}
p(\rho)=p_t+\frac{1}{3}(p_r-p_t),\label{eq4a}
\end{equation}
The relation \( \frac{1}{3} = \langle \delta_i{}^1 \delta^j{}_1 \rangle \) is utilized, and considering the equation of state (EOS) \( p = \omega \rho \) with \(-1 < \omega < -\frac{1}{3}\) for quintessence matter, the expression \( p_t = \frac{1}{2}(1+3\omega)\rho \) is derived from Eq.~(\ref{eq4a}). This relationship is consistent with the radial pressure \( p_r = -\rho \) as discussed in~\cite{jim2}. In our scenario, the non-linear EOS for the Chaplygin Dark Fluid (CDF) is given by \( p = -\frac{B}{\rho} \), where \( B \) is a positive constant. This leads to the tangential pressure \( p_t = \frac{1}{2}\rho - \frac{3B}{2\rho} \) with \( p_r = -\rho \). Consequently, the stress-energy tensor of the CDF can be expressed as.
\begin{eqnarray}
&-\rho={{T}_t}^t={{T}_r}^r, \label{Ttr}~\\
&\frac{1}{2}\rho-\frac{3B}{2\rho}={{T}_{\theta}}^{\theta}={{T}_{\phi}}^{\phi}.~\label{Tangular}
\end{eqnarray}
On the cosmological scale, the anisotropy of the Cosmic Dust Fluid (CDF) diminishes, leading to an equation of state (EOS) where pressure (\(p\)) is expressed as \(-B/\rho\). Additionally, a crucial condition, \(T_t^t = T_r^r\), must be satisfied, ensuring that the metric components \(g(r)\) and \(f(r)\) are related through an appropriate time re-scaling, preserving generality. The calculation of the Einstein tensor components is subsequently performed as
\begin{eqnarray}
&{{G}_r}^r=\frac{1}{r^2}(f+rf'-1)={{G}_t}^t,~\label{Gtr}\\
&\frac{1}{2r}(2f'+rf'')={{G}_{\theta}}^{\theta}={{G}_{\phi}}^{\phi}. \label{Gthetafai}
\end{eqnarray}
Now, by using the Eqs.~(\ref{Ttr})-(\ref{Tangular}) and Eqs.~(\ref{Gtr})--(\ref{Gthetafai}), one can get the following relation
\begin{eqnarray}
&-\rho=\frac{1}{r^2}(f+rf'-1)+\Lambda,~\label{graviequationsrt}\\
&\frac{1}{2}\rho-\frac{3B}{2\rho}=\frac{1}{2r}(2f'+rf'')+\Lambda.~\label{graviequationsthetafai}
\end{eqnarray}
We now examine the existence of the cosmological constant. The two above differential equations allow us to calculate the two unknown functions, i.e., $f(r)$ and $\rho(r)$, analytically. Now, one can simply the energy density of CDF by solving the above set of differential Eqs. (\ref{graviequationsrt})--(\ref{graviequationsthetafai}) as:
\begin{equation}
\rho(r)=\sqrt{B+\frac{q^2}{r^6}}, \label{CDFenergydensity}
\end{equation}
where a normalization factor $q>0$ denotes the CDF's intensity. Additionally, the conservation law for the stress-energy tensor $\nabla_{b}T^{ab}=0$ directly yields Eq.~(\ref{CDFenergydensity}). It is evident that the CDF energy density can be approximated by $r^6\ll q^2/B$ as
\begin{equation}
\rho(r)\approx\frac{q}{r^3}, \label{CDFenergydensitysmallr}
\end{equation}
demonstrating that the CDF behaves as though its energy density were dependent on $r^{-3}$ for matter content. Now, by using Eq.~(\ref{CDFenergydensity}) into Eq.~(\ref{graviequationsrt}), we have the final analytical solution for $f(r)$ ~\cite{rjim6100,jim6} as
\begin{equation}
f(r)=1-\frac{2M}{r}+\frac{q}{3r}{\rm ArcSinh}\frac{q}{\sqrt{B}r^3}-\frac{r^2}{3}\sqrt{B+\frac{q^2}{r^6}}-\frac{ r^2}{3}\Lambda, 
\label{14}
\end{equation}
where $M$ is the BH mass, however in the current case, the BH mass is regarded as a point mass BH.\\\\
From the equation of horizon $f(r)=0$ and pressure $P=-\frac{\Lambda}{8\pi}$ \cite{J1,J2,J3}, we can deduce the relation between the BH mass $M$ and its event horizon radius $r_h$, expression as follows
\begin{equation}  \label{m1}
\begin{aligned}
    M = \frac{1}{6} & \left(r^3 \left(-\sqrt{B+\frac{q^2}{r^6}}\right) + q \sinh^{-1}\left(\frac{q}{\sqrt{B} r^3}\right) \right. \\
    & \left. - 8 \pi P r^3 + 3 r\right)
\end{aligned}
\end{equation}
The Hawking temperature of BH related to surface gravity can be obtained  as
\begin{equation} \label{t1}
\begin{aligned}
    T & = - \Bigg \{\frac{2 r^6 \sqrt{B+\frac{q^2}{r^6}}-\frac{3 q^2}{\sqrt{B+\frac{q^2}{r^6}}}+\frac{3 q^2}{\sqrt{B} \sqrt{\frac{q^2}{B r^6}+1}}}{12 \pi  r^5}\\
    & +\frac{\frac{3 q^2}{\sqrt{B} \sqrt{\frac{q^2}{B r^6}+1}}+q r^3 \sinh ^{-1}\left(\frac{q}{\sqrt{B} r^3}\right)-6 M r^3+16 \pi  P r^6}{12 \pi  r^5} \Bigg \}
\end{aligned}
\end{equation}
The BH entropy with the help of area law is defined as  \cite{J3,J4,J5}
\begin{equation}\label{B4}
S=\pi r^2.
\end{equation}
The volume parameter is  defined as
\begin{equation}\label{Ba4}
V=\bigg(\frac{\partial M}{\partial P}\bigg)_{S,q_m},
\end{equation}
calculated as
\begin{equation}\label{B5}
V=\frac{r^3}{6}
\end{equation}
To find  more data about a phase transition, we study thermodynamic a quantity such as heat capacity. By applying the  standard definition of heat capacity follows as 
\begin{equation}\label{B5}
\text{Cp}=\frac{\partial S}{\partial T}=-\frac{8 \pi ^2 r}{\frac{\left(B r^6-2 q^2\right) \sqrt{B+\frac{q^2}{r^6}}}{B r^6+q^2}+8 \pi  P+\frac{1}{r^2}}.
\end{equation}
The coordinate location of the horizons in this spacetime when using Schwarzschild coordinates ($t$,$r$,$\theta$,$\phi$) is given by imposing $g^{rr} = 0$. Since the analytical expression of the lapse function Eq. (\ref{14}) is a function of multiple variables, it can be plotted against the radial coordinate to find the locations of horizons in the spacetime geometry. Such a curve can be seen in Fig. (\ref{fig:1}). The coordinate location of the horizons for different values of $q, B$ and $\Lambda$ is provided in Tab. (\ref{table1}). It can be seen from Fig. (\ref{fig:1}) that the lapse function curve crosses the horizontal axis at two points suggesting the presence of two horizons i.e. Cauchy (inner horizon) and the event horizon.  Also, by studying the radial null curves in this geometry i.e. setting $ds^{2} = 0$, $d\theta = d\phi =0$ we get:
\begin{equation}
\begin{aligned}
    \frac{dr}{dt}& = \pm \bigg (1-\frac{1}{3} r^2 \sqrt{B+\frac{q^2}{r^6}}+\frac{q \sinh ^{-1}\left(\frac{q}{\sqrt{B} r^3}\right)}{3 r} \\
    &-\frac{2 M}{r}-\frac{\Lambda  r^2}{3} \bigg ).
\end{aligned}
\end{equation}
This equation also defines the coordinate speed of light in this metric. As the radial coordinate approaches the values provided in Tab. (\ref{table1}), $\frac{dr}{dt} \rightarrow 0$.
\begin{table} \label{T1}
\begin{center}
\begin{tabular}{|c|c|c|}
\hline
\multicolumn{3}{|c|}{Location of horizons for AdS black hole spacetime} \\
\hline
Spacetime parameters & Cauchy horizon & Event horizon \\
\hline
$q = 0.8$, $B = 0.4$, $\Lambda = -0.70$ &0.089&1.833\\
\hline
$q = 0.8$, $B = 0.7$, $\Lambda = -0.90$ &0.080&1.846\\
\hline
$q = 0.8$, $B = 0.9$, $\Lambda = -1$ &0.077&1.870\\
\hline
\end{tabular}
\caption{Location of event and Cauchy horizons for AdS black hole spacetime. We take $M=1$ for our analysis} \label{table1}
\end{center}
\end{table}
\newpage
\section{Joule-Thomson Expansion}~\label{sec3}
One of the most well-known and classical physical processes to explain the change in the temperature of gas from a high-pressure section to reduced pressure
through a porous plug is called  Joule-Thomson expansion. The main focus is on the gas expansion process, which expresses the cold effect (when the temperature drops) and the heat effect (when the temperature increases), with the enthalpy remaining constant throughout the process. This change depends upon the coefficient of Joule-Thomson as   \cite{J1,J2,J3,J4,J5,J6}
\begin{equation}\label{j1}
\mu_{JT}=\bigg(\frac{\partial T}{\partial
P}\bigg)_{H}=\frac{1}{C_{p}}\bigg[T\bigg(\frac{\partial V}{\partial
T}\bigg)_{p}-V\bigg].
\end{equation}
\begin{figure}
\includegraphics[width= 0.99\linewidth]{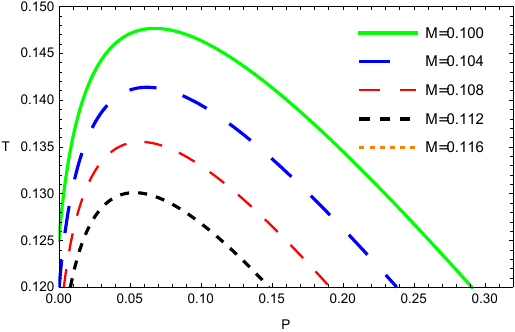}
\caption{Plot of inversion curves in  $T-P$ plane  with fixed values of $B=0.001$ and $q=0.04$.}\label{fj1} 
\end{figure}
\begin{figure}
\includegraphics[width= 0.99\linewidth]{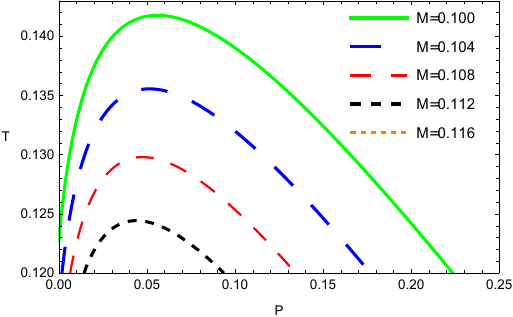}
\caption{Plot of inversion curves in  $T-P$ plane  with fixed values of $B=0.001$ and $q=0.05$.}\label{fj2} 
\end{figure}
\begin{figure}
\includegraphics[width= 0.99\linewidth]{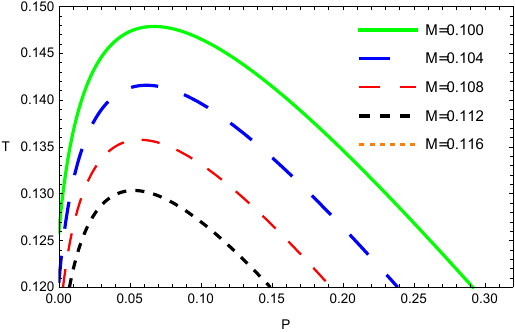}
\caption{Plot of inversion curves in  $T-P$ plane  with fixed values of $B=0.05$ and $q=0.04$.}\label{fj3} 
\end{figure}
\begin{figure}
\includegraphics[width= 0.99\linewidth]{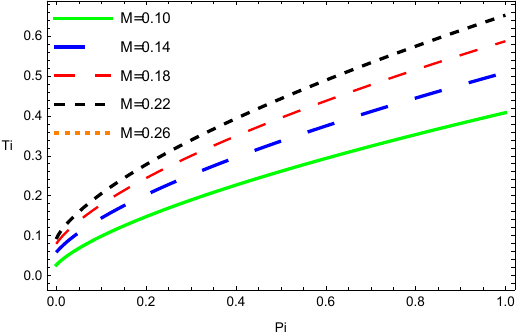}
\caption{Plot of isenthalpic curves in  $Ti-Pi$ plane  with fixed values of $B=0.03$ and $q=0.04$. }\label{fj4} 
\end{figure}
\begin{figure}
\includegraphics[width= 0.99\linewidth]{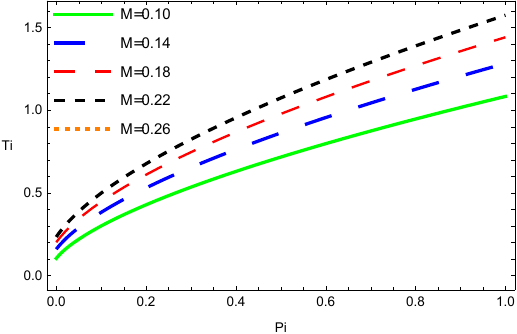}
\caption{Plot of isenthalpic curves in $Ti-Pi$ plane with fixed values of $B=0.03$ and $q=0.08$.}\label{fj5} 
\end{figure}
\begin{figure}
\includegraphics[width= 0.99\linewidth]{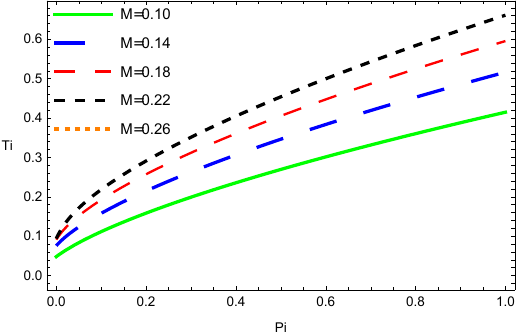}
\caption{Plot of isenthalpic curves in  $Ti-Pi$ plane  with fixed values of $B=0.06$ and $q=0.04$. }\label{fj6} 
\end{figure}
Putting the  basic quantities in Eq. (\ref{j1}) coefficient calculated as
\begin{equation} \label{j2}
\begin{aligned}
    \mu_{JT} & = \Bigg \{\frac{q^2 \left(-5 r^2 \sqrt{B+\frac{q^2}{r^6}}-16 \pi  P r^2+4\right)}{48 \pi ^2 \left(B r^6+q^2\right)} \\
    & \frac{-2 B r^6 \left(r^2 \sqrt{B+\frac{q^2}{r^6}}+8 \pi  P r^2-2\right)}{48 \pi ^2 \left(B r^6+q^2\right)}    \Bigg \}
\end{aligned}
\end{equation}
It is clear from a comparison of these figures that the zero point of the Hawking temperature and the divergence point of the coefficient of
Joule-Thomson is the same. This point of divergence gives information on the Hawking temperature and corresponds to the most extreme BHs. From Eq. (\ref{j2}) utilizing the well known condition $\mu_{JT}=0$, the temperature inversion occurs as\\\\
\begin{equation}\label{j3}
\begin{aligned}
    T_i & = \Bigg \{\frac{q^2 \left(2 r^2 \sqrt{B+\frac{q^2}{r^6}}-8 \pi  P r^2-1\right)}{12 \pi  r \left(B r^6+q^2\right)}\\
    & \frac{-B r^6 \left(r^2 \sqrt{B+\frac{q^2}{r^6}}+8 \pi  P r^2+1\right)}{12 \pi  r \left(B r^6+q^2\right)}\Bigg \}
\end{aligned}
\end{equation}
From Figs. (\ref{fj1})-(\ref{fj3}), the inversion temperature increases with variations of important parameters respectively. Compared with the van der Waals fluids,  the above results in $Ti-Pi$-plan at low pressure, the inversion temperature $T_i$ decreases with the increase of $P$, which shows the opposite behavior for higher pressure. It is also clear that, unlike the case with van der Waals fluids. We show the constant mass curves for various values of the nonlinear parameter $k$ and the massive gravity couplings  From this figure, we observe that with the same mass, as the nonlinear parameter and the massive gravity couplings c1,2 increase, the constant mass curves tend to shrink towards the lower pressure and temperature. In Figs. (\ref{fj4})-(\ref{fj6}),  we study isenthalpic curves ($Ti-P_i$- plane) by assuming different values of BH mass which are investigated in  Eq. (\ref{j3}) with a larger root of $r_h$. We show the isenthalpic curves and the inversion curves of BH  and this result is consistent \cite{J5,J6}. Heating and cooling zones are characterized by the inversion curve, and isenthalpic curves possess positive slopes above the inversion curve. In contrast, the pressure always falls in a Joule-Thomson expansion and the slope changes sign when heating happens below the inversion curve. The heating process appears at higher temperatures, as indicated by the negative slope of the constant mass curves in the Joule-Thomson expansion.
\section{RADIAL GEODESICS}~\label{sec4}
In this section we will study about the radial geodesics in AdS black hole spacetime exhibiting dark fluid with Chaplygin like equation of state. Radial geodesics are such that $\dot{\theta}$ and $\dot{\phi}$ = $0$. For our analysis, we take the geodesics to be in the equatorial plane s.t. $\theta = \frac{\pi}{2}$. Any general spherically symmetric static spacetime can be written as:
\begin{equation} 
    ds^{2} = -g_{tt} dt^{2} + g_{rr} dr^{2} + r^{2} (d\theta^{2} + \sin^{2}\theta d\phi^{2})
    \label{25}
\end{equation}
where $g_{tt}$ and $g_{rr}$ are the components of the metric tensor, and the azimuthal part shows the spherical symmetry of the spacetime. For any freely falling particle in this spacetime the energy ($E$) and the angular momentum ($L$) per unit rest mass is always conserved. It is a direct consequence of the spherical and temporal symmetry of the spacetime \cite{Duv1}. The conserved quantities can be written as:
\begin{equation} 
E = g_{tt} \frac{dt}{d\tau} ;\quad L = r^{2} \sin^{2}\theta \frac{d\phi}{d\tau}
    \label{26}
\end{equation}
where $\tau$ is the proper time of the particle. A freely falling particle always follow a timelike geodesic s.t the $u^{\mu} u_{\mu} = -1$, where $u^{\mu}$ is the particle's four velocity. From this normalization of four velocity, we can obtain the radial equation of motion for a freely falling particle in the equatorial plane, which can be written as:
\begin{equation} 
  \frac{\dot{r}^{2}}{2} = \frac{E^{2}-f(r)}{2}
  \label{27}
\end{equation}
For a radially in-falling particle from rest at a point $r=b$, the energy of the particle can be given as $E = \sqrt{f(r=b)}$, since $\frac{dr}{d\tau} = 0$ at $r=b$. We can also define the Newtonian radial acceleration by differentiating Eq. (\ref{27}). w.r.t proper time. This gives us the expression:
\begin{equation} 
    A^{r} = \frac{-f'(r)}{2}
    \label{28}
\end{equation}
For the purpose of this study, inserting the derivative of lapse function in Eq. (\ref{28}), we get the Newtonian acceleration as:

\begin{equation}
\begin{aligned}
     A^{r}& = \Bigg \{\frac{ \dfrac{2\sqrt{\frac{q^2}{r^6}+B}\,r}{3}+\dfrac{2{\Lambda}r}{3}+\dfrac{q\operatorname{arcsinh}\left(\frac{q}{\sqrt{B}\,r^3}\right)}{3r^2}}{2}  \\
    & \frac{ -\dfrac{2M}{r^2}+\dfrac{q^2}{\sqrt{B}\sqrt{\frac{q^2}{Br^6}+1}\,r^5}-\dfrac{q^2}{\sqrt{\frac{q^2}{r^6}+B}\,r^5} }{2}\Bigg \}
\end{aligned}
\end{equation}
Since, in this spacetime the value of $\Lambda < 0$, we can infer that negative values of the cosmological constant provides increased attraction to the black hole. Also, any test particle falling freely from rest from a point $r=b > R_{eh}$ can turn back after reaching a point $R^{stop}$. This point can be calculated as the root of the equation $E^{2} = f(r)$. 
\section{TIDAL FORCE ANALYSIS}~\label{sec5}
We shall dig into the intricacies of the equations regarding tidal forces in our spacetime in this part. We utilise the following equation for the spacelike components of the geodesic deviation vector $\eta^{\mu}$,  to analyse the equation for the distance between two infinitesimally close and free falling particles. 
\begin{eqnarray}
    \frac{D^{2} \eta^{\mu}}{D\tau^{2}} = R^{\mu}_{\alpha \beta \gamma} v^{\alpha}  v^{\beta} \eta^{\gamma},
    \label{30}
\end{eqnarray}
where $R^{\mu}_{\alpha \beta \gamma}$ is the Riemann curvature tensor and $v^{\mu}$ is the unit tangent vector to the geodesic. A test body travelling in space alone under the influence of gravity is known to follow a geodesic \cite{Duv2}.  since of the variation in curvature at each point on the geodesic, and since each location on the test body tends to follow a different geodesic, the body experiences a change in acceleration, resulting in a stretching and squeezing action known as tidal forces. This is why the Riemann curvature tensor is employed to examine gravity-induced tidal interactions.\\\\
We utilise the tetrad formalism to compute the tidal forces in the falling body's frame \cite{Duv3}. Tetrads or vierbien are geometric objects that create a set of local coordinate bases, i.e. a locally defined collection of four linearly independent vector fields. A tetrad frame establishes a local inertial reference frame at each point on a geodesic, where the equations of special relativity apply. Lorentz transformations cannot provide adequate information on the metric link because they can connect an endless number of orthonormal bases at a particular place. To fix this, we apply the orthonormal basis to coordinate basis transformation:
\begin{eqnarray}
    \overrightarrow{e}_{\mu} = \hat{e}^{\hat{i}}_\mu  \overrightarrow{e}_{\hat{i}},
    \label{31}
\end{eqnarray}
where $\overrightarrow{e}_{\mu}$ represents the coordinate basis, $\overrightarrow{e}_{\hat{i}}$ represents the orthonormal basis and $\hat{e}^{\hat{i}}_\mu $ are the tetrad components. The metric tensor components in the tetrad basis are given as \cite{Duv4}:
\begin{eqnarray}
    g_{\mu\nu} = \eta_{\hat{i}\hat{j}} \hat{e}^{\hat{i}}_\mu \hat{e}^{\hat{\hat{j}}}_\nu. 
    \label{32}
\end{eqnarray}
This is the crucial equation for using orthonormal basis in curved spacetime. Tetrad components for freely falling frames are available from Eq. (\ref{31}) and Eq. (\ref{32}).
\begin{equation}
    \hat{e}^{\mu}_{\hat{0} } = \bigg\{\frac{E}{f(r)} , -\sqrt{E^{2}-f(r)}, 0, 0\bigg\} ,
    \label{33}
\end{equation}
\begin{equation}
    \hat{e}^{\mu}_{\hat{1} } = \bigg\{\frac{-\sqrt{E^{2}-f(r)}}{f(r)}, E, 0, 0\bigg\},
    \label{34}
\end{equation}
\begin{equation}
    \hat{e}^{\mu}_{\hat{2} } = \bigg\{0, 0, \frac{1}{r}, 0\bigg\}, 
    \label{35}
\end{equation}
\begin{equation}
    \hat{e}^{\mu}_{\hat{3} } = \bigg\{0 ,0, 0, \frac{1}{r\sin\theta}\bigg\}, 
    \label{36}
\end{equation} 
where the tetrad components follow the following rule:
\begin{equation}
     \hat{e}^{\mu}_{\hat{i} }  \hat{e}^{\hat{j}}_{\mu}  = \delta^{\hat{j}}_{\hat{i}}.
     \label{37}
\end{equation}
We can see that the component $\hat{e}^{\mu}_{\hat{0} } = v^{\mu}$ is the tangent vector to the geodesic, i.e. the particle's four velocity. Furthermore, the geodesic deviation vector adheres to the transformation rule from global to local orthonormal coordinates.
\begin{equation}
    \eta^{\mu} = \hat{e}^{\mu}_{\hat{i} } \eta^{\hat{i}}.
    \label{38}
\end{equation}
For spherically symmetric spacetimes, including EBR black hole spacetime, the non-vanishing independent components of the Riemann tensor are given by \cite{Duv5}:
\begin{eqnarray}
   R^{1}_{2 1 2} = \frac{-r f'(r)}{2} , 
   \label{39}
\end{eqnarray}
\begin{eqnarray}
    R^{1}_{0 1 0} = \frac{f(r) f''(r)}{2},
    \label{40}
\end{eqnarray}
\begin{equation}
    R^{1}_{3 1 3} = \frac{-r f'(r)\sin^{2}\theta}{2} ,
    \label{41}
\end{equation}
\begin{eqnarray}
    R^{2}_{0 2 0} = R^{3}_{0 3 0} = \frac{f(r) f'(r)}{2r},
    \label{42}
\end{eqnarray}
\begin{eqnarray}
    R^{2}_{3 2 3} = \sin^{2}\theta \bigg (1 - r^{2} f(r) \bigg ) 
    \label{43}
\end{eqnarray}
We utilise the tetrad formalism to compute the Riemann curvature tensor components in tetrad basis as follows:
\begin{equation}
    R^{\hat{i}}_{\hat{j} \hat{k} \hat{l}} = R^{\mu}_{\alpha \beta \gamma} e^{\hat{i}}_\mu e^{\alpha}_{\hat{j}} e^{\beta}_{\hat{k}} e^{\gamma}_{\hat{l}}.
     \label{44}
\end{equation}
Following the Eq. (\ref{44}), the tidal tensor components are given by:  
\begin{eqnarray}
    R^{\hat{1}}_{\hat{0} \hat{1} \hat{0}} = \frac{f''(r)}{2},
    \label{45}
\end{eqnarray}
\begin{eqnarray}
   R^{\hat{2}}_{\hat{0} \hat{2} \hat{0}} =  R^{\hat{3}}_{\hat{0} \hat{3} \hat{0}} =  \frac{f'(r)}{2r},
   \label{46}
\end{eqnarray}

\subsection{TIDAL FORCE EQUATIONS}\label{III A}
Upon obtaining the expressions from Eq. (\ref{45}-\ref{46}), we can obtain the relative acceleration between two nearby particles as:
\begin{eqnarray}
   \frac{D^{2}\eta^{\hat{r}}}{D\tau^{2}} =  \frac{-f''(r)}{2} \eta^{\hat{r}},
   \label{47}
\end{eqnarray}
\begin{eqnarray}
    \frac{D^{2}\eta^{\hat{\theta}}}{D\tau^{2}} = -\frac{f'(r)}{2r}  \eta^{\hat{\theta}},
    \label{48}
\end{eqnarray}
\begin{eqnarray}
    \frac{D^{2}\eta^{\hat{\phi}}}{D\tau^{2}} = -\frac{f'(r)}{2r} \eta^{\hat{\phi}}.
    \label{49}
\end{eqnarray}
By substituting the values of $f(r)$ and its higher derivatives in Eq. (\ref{47}-\ref{49}) we get:
\begin{widetext}
\begin{equation}
\begin{aligned}
\frac{D^{2}\eta^{\hat{r}}}{D\tau^{2}} & = \Bigg \{\frac{\dfrac{-2q\operatorname{arcsinh}\left(\frac{q}{\sqrt{B}\,r^3}\right)}{3r^3}+\dfrac{4M}{r^3}-\dfrac{6q^2}{\sqrt{B}\sqrt{\frac{q^2}{Br^6}+1}\,r^6}+\dfrac{3q^2}{\sqrt{\frac{q^2}{r^6}+B}\,r^6}+\dfrac{3q^4}{B^\frac{3}{2}\cdot\left(\frac{q^2}{Br^6}+1\right)^\frac{3}{2}r^{12}}}{2} \\ &   \frac{-\dfrac{3q^4}{\left(\frac{q^2}{r^6}+B\right)^\frac{3}{2}r^{12}}+ \dfrac{2\sqrt{\frac{q^2}{r^6}+B}}{3}+\dfrac{2{\Lambda}}{3}}{2} \Bigg \} \eta^{\hat r},\label{50}
\end{aligned}
\end{equation}
\begin{eqnarray}
\frac{D^{2}\eta^{\hat{\theta}}}{D\tau^{2}}= \frac{ \dfrac{2\sqrt{\frac{q^2}{r^6}+B}\,r}{3}+\dfrac{2{\Lambda}r}{3}+\dfrac{q\operatorname{arcsinh}\left(\frac{q}{\sqrt{B}\,r^3}\right)}{3r^2}-\dfrac{2M}{r^2}+\dfrac{q^2}{\sqrt{B}\sqrt{\frac{q^2}{Br^6}+1}\,r^5}-\dfrac{q^2}{\sqrt{\frac{q^2}{r^6}+B}\,r^5} }{2r}\eta^{\hat{\theta}},
    \label{51}
\end{eqnarray}
\begin{eqnarray}
    \frac{D^{2}\eta^{\hat{\phi}}}{D\tau^{2}} = \frac{ \dfrac{2\sqrt{\frac{q^2}{r^6}+B}\,r}{3}+\dfrac{2{\Lambda}r}{3}+\dfrac{q\operatorname{arcsinh}\left(\frac{q}{\sqrt{B}\,r^3}\right)}{3r^2}-\dfrac{2M}{r^2}+\dfrac{q^2}{\sqrt{B}\sqrt{\frac{q^2}{Br^6}+1}\,r^5}-\dfrac{q^2}{\sqrt{\frac{q^2}{r^6}+B}\,r^5} }{2r} \eta^{\hat{\phi}}.
    \label{52}
\end{eqnarray}
\end{widetext}
\section{RADIAL TIDAL FORCE}~~\label{sec6}
\begin{figure*}[htbp]
    \centering
    \includegraphics[width = 0.45\linewidth]{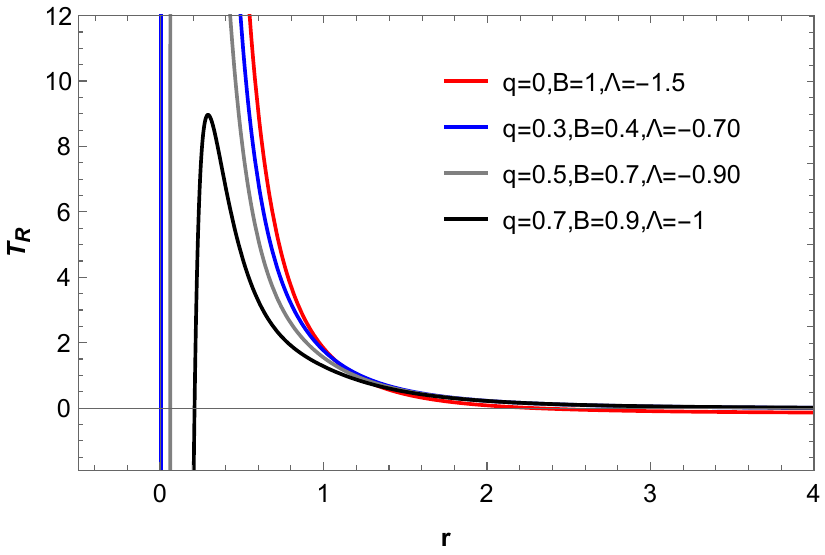}
    \qquad
    \includegraphics[width = 0.45\linewidth]{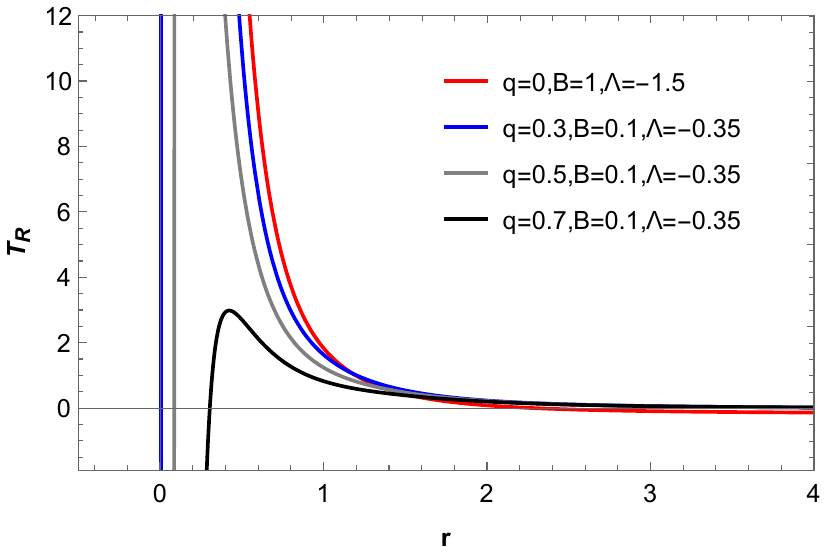}
    \caption{On left, we plot the radial tidal force profile for varying values of $q, B, \Lambda$. On right, we plot the force profile for different values of $q$ while keeping $B$ and $\Lambda$ fixed. It can be observed that a small drop in the value of $q$ causes a sharp increase in the peak value of the force. We use $M=1$.}
    \label{fig:9}
\end{figure*}

In Fig. (\ref{fig:8}), we plot the radial tidal force for AdS black hole for fixed value of $q$ and changing values of $B$ and $\Lambda$. It can be seen that the presence of non-zero values of $q$ changes the monotonicity of the dependency of the radial tidal force on the radial coordinate $r$. Fig. (\ref{fig:9}) shows the radial force profile when the value of $q$ is varied, and the values of $B$ and $\Lambda$ are fixed. Fig. (\ref{fig:9}) also takes into account different values for $q$,$B$ and $\Lambda$ respectively. It can be seen that in all the three cases, the asymptotic value of radial tidal force is non-zero. At infinitely large distances where the energy density of the cold dark fluid can be approximated by $\rho(r) = \sqrt{B}$, the radial tidal force can be given by the equation:
\begin{equation}
  \lim _{r \to \infty }\frac{D^{2}\eta^{\hat{r}}}{D\tau^{2}} = \frac{1(\sqrt{B}+\Lambda)}{3} \eta^{\hat r}
  \label{53}
\end{equation}
This proves that the radial tidal force at $r \rightarrow \infty$ is solely governed by the presence of an effective non-zero cosmological constant. In all the three cases, the radial tidal forces increase drastically at short distances i.e. when the in-falling particle approaches the central singularity. The monotonically increasing trend or the radial stretching switches its behavior at a certain value of radial coordinate, after which it falls rapidly to become compressive. The transition from radial stretching to compression happens at a crossover radial coordinate, where the tidal force is momentarily zero, after which it approaches infinity as $r \rightarrow 0$. The analytic expression for it can be computed by equating the RHS of Eq. (\ref{47}) to zero:
\begin{equation}
    R_0^{rtf} = \frac{\frac{4M}{r^{3}} - g''(r)}{2} = 0
    \label{54}
\end{equation}
where the function $g(r)$ is taken for simplicity in calculation and representation, and is given by:
\begin{eqnarray}
    g(r) = -\frac{1}{3} r^2 \sqrt{B+\frac{q^2}{r^6}}+\frac{q \sinh ^{-1}\left(\frac{q}{\sqrt{B} r^3}\right)}{3 r}-\frac{\Lambda  r^2}{3}.
    \label{55}
\end{eqnarray}
In all the cases, the radial tidal force profile peaks at a certain value of the radial coordinate $r$. The peak behavior is different for changing values of $B$ and $\Lambda$. From Fig. (\ref{fig:8}) it can be seen that for $q=0.8$, the maximum value of radial stretching is attained for $B=0.9$ and $\Lambda = -1$. The peak value declines as the value of $B$ is reduced or as the value of $\Lambda$ is increased. An opposite trend can be seen from Fig. (\ref{fig:9}). It can be inferred that for fixed values of $B$ and $\Lambda$, the peak value of radial stretching drops as the value of $q$ is increased. It can also be seen that the peak value is achieved inside the event horizon and before the Cauchy horizon of the black hole i.e. $R_{ch} < R_{peak}^{rtf}< R_{eh}$ where the $R_{ch}$ and $R_{eh}$ denotes the Cauchy horizon and event horizon respectively. For eg. for the parameters $q=0.8, B=0.4, \Lambda = -0.70$, the peak value is attained at approximately $r = 0.50$ which lies inside the event horizon located at $r=1.833$ and outside the Cauchy horizon located at $r=0.089$. The analytic expression for peak value can be given by equating the first derivative of Eq. (\ref{47}) to zero. Since the tidal force expressions involve the use of higher order polynomials, a more general expression for peak value can be written as:
\begin{eqnarray}
   R_{peak}^{rtf} =  \frac{d}{dr} \bigg(\frac{\frac{4M}{r^{3}} - g''(r)}{2} \bigg) =0.
   \label{56}
\end{eqnarray}
The graphs for different values of the spacetime parameters are shown below:
\begin{figure}
\includegraphics[width= 0.99\linewidth]{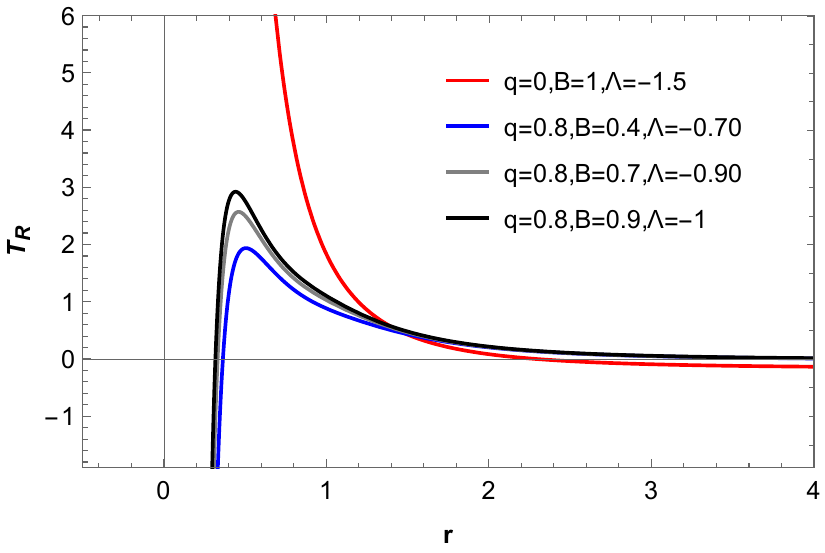}
\caption{Radial tidal force as a function of the radial coordinate for fixed value of $q=0.8$. It can be observed that the force changes monotonicity and vanishes at a particular radial coordinate which lies between the Cauchy horizon and the event horizon of the black hole for non-zero values of $q$.We take $M=1$}\label{fig:8} 
\end{figure}
\subsection{ANGULAR TIDAL FORCE}\label{III C}
\begin{figure*}[htbp]
    \centering
    \includegraphics[width = 0.45\linewidth]{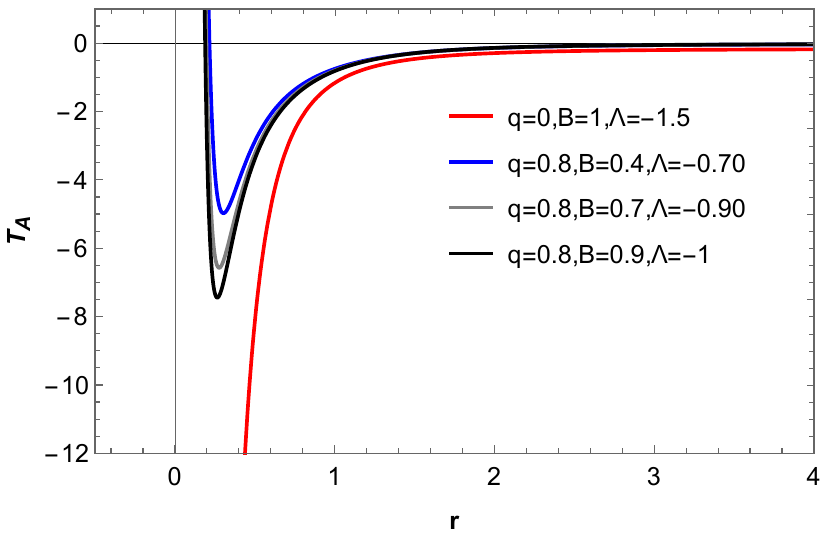}
    \caption{Angular tidal force as a function of the radial coordinate for fixed value of $q=0.8$. It can be observed that the force changes monotonicity and vanishes at a particular radial coordinate which lies between the Cauchy horizon and the event horizon of the black hole for non-zero values of $q$.We take $M=1$}
    \label{fig:10}
\end{figure*}

\begin{figure*}[htbp]
    \centering
    \includegraphics[width = 0.45\linewidth]{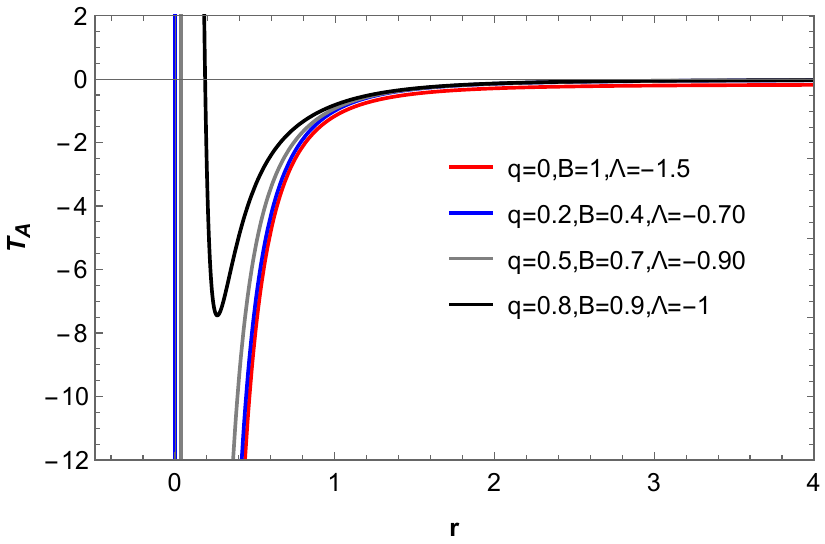}
    \qquad
    \includegraphics[width = 0.45\linewidth]{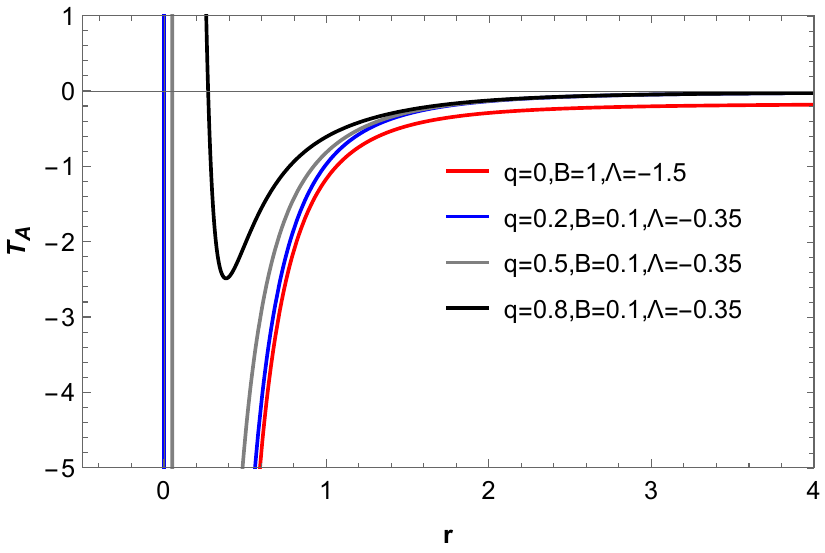}
    \caption{On left, we plot the angular tidal force profile for varying values of $q, B, \Lambda$. On right, we plot the force profile for different values of $q$ while keeping $B$ and $\Lambda$ fixed. It can be observed that a small drop in the value of $q$ causes a drastic decrease in the compressiveness of the tidal force. We use $M=1$.}
    \label{fig:11}
\end{figure*}

The angular tidal forces exhibit compressive nature unlike the radial tidal forces as a radially in-falling body approaches the central singularity. But, similar to the radial case, angular tidal forces also reach a peak compressive value and then rapidly fall to become momentarily zero at a particular value of radial coordinate $r$ and switching their behavior from compressive to radial stretching. Fig. (\ref{fig:10}) portray the angular force profile w.r.t the radial coordinate $r$, keeping the value of $q$ fixed and varying the values of $B$ and $\Lambda$. We use $q=0.8$ for the analysis.\\\\
It can be seen that for non-zero values of $q$, the monotonicity of the force profile changes. Fig. (\ref{fig:11}) show the trend for different values of $q$ and also displays the trend for different values of $q$, $B$ and $\Lambda$. In all the cases the asymptotic value of the angular tidal force is non-zero, and is given in the limit of $r \rightarrow \infty$ by the expression
\begin{equation}
  \lim _{r \to \infty }\frac{D^{2}\eta^{\hat{i}}}{D\tau^{2}} = \frac{1(\sqrt{B}+\Lambda)}{3} \eta^{\hat i}
  \label{57}
\end{equation}
where $i = \{\theta,\phi\}$. Similar to the case of radial tidal forces, the values of angular tidal forces at infinitely large distances is solely governed by the presence of a non-zero effective cosmological constant. The crossing radial coordinate at which the angular tidal force is zero is given by equating the R.H.S of Eq. (\ref{48}) to zero as:
\begin{eqnarray}
R_0^{atf}=\frac{\frac{2M}{r^{2}}+g'(r)}{2r} =0. 
\label{58}
\end{eqnarray}
where the term $g(r)$ is given by Eq. (\ref{55}). After becoming radially stretching in nature, the angular tidal forces continue to grow till infinity in the limit $r \rightarrow 0$. Following the convention in Eq. (\ref{55}), the peak value of angular tidal force can be calculated by equating the first derivative of the expression in Eq. (\ref{48}) to zero.
\begin{eqnarray}
    R_{peak}^{atf} = \frac{d}{dr} \bigg (\frac{\frac{-2M}{r^{2}} - g'(r)}{2r} \bigg ) =0.
    \label{59}
\end{eqnarray}
From Fig. (\ref{fig:10}), it can be seen that for a fixed value of $q=0.8$, the peak value of angular tidal force goes on decreasing as the value of $B$ is increased. For instance the black curve shows minimum compression for the values of $B=0.9, \Lambda=-1$ in contrast to the blue curve which shows maximum compression for values of $B=0.4, \Lambda = -0.70$. An opposite behavior can be seen from Fig. (\ref{fig:11}), where as the value of $q$ is increased while keeping $B$ and $\Lambda$ fixed, the angular compression also increases. Similar to the radial case, the peak value is achieved inside the event horizon but before the Cauchy horizon i.e. $R_{ch} < R_{peak}^{atf}< R_{eh}$. For eg. when $q=0.5, B=0.7, \Lambda = -0.90$, the Cauchy horizon and the event horizon lies at $r = 0.0143$ and $r = 1.8570$ respectively. The peak value as observed from Fig. (\ref{fig:10}) is approximately at $r=0.19$ which lies inside the two horizons. The graphs for angular tidal force profiles for different values of the spacetime parameters are shown below.

\newpage
\section{GEODESIC DEVIATION}~\label{sec7}
In this section we discuss the variation of geodesic deviation vector with the radial coordinate in SBR black hole spacetime. The deviation vector measures the deformation of a body falling radially in any spacetime geometry. We can convert Eq. (\ref{47}) and (\ref{48}) in second derivatives w.r.t. $r$ by substituting $dr/d\tau$ = $-\sqrt{E^{2}-f(r)}$ which results from Eq. (\ref{33}). This gives us the following second order differential equations in $r$:
\begin{eqnarray}
  ( E^2-f(r)) \frac{D^{2}\eta^{\hat{r}}}{D r^{2}} - \frac{f'(r)}{2} \frac{D \eta^{\hat{r}}}{dr}+\frac{f''(r)}{2} \eta^{\hat{r}} = 0,
  \label{60}
\end{eqnarray}
\begin{eqnarray}
    ( E^2-f(r)) \frac{D^{2}\eta^{\hat{i}}}{D r^{2}} - \frac{f'(r)}{2} \frac{D \eta^{\hat{i}}}{dr} + \frac{f'(r)}{2r} \eta^{\hat{i}} =0.
    \label{61}
\end{eqnarray}
where $i = \{\theta,\phi\}$. The analytic solution for Eq. (\ref{60}) and (\ref{61}) as pointed out in \cite{crispino2016tidal} can be given as:
\begin{eqnarray}
  \eta^{\hat{r}} (r) = \sqrt{E^{2}-f(r)}\bigg[C_{1}+C_{2}\int \frac{dr}{(E^{2}-f(r))^{3/2}}\bigg] ,
  \label{62}
\end{eqnarray}
\begin{eqnarray}
  \eta^{\hat{i}} (r) = \bigg [C_{3} +C_{4} \int \frac{dr}{r^{2} \sqrt{(E^{2}-f(r))}} \bigg ] r.
  \label{63}
\end{eqnarray}
where $C_{1}, C_{2}, C_{3}, C_{4}$ are constants of integration. In order to find their value, we numerically solve the differential equations by imposing some initial conditions. For the purpose of this study we take the following initial conditions.
\begin{eqnarray}
    \eta^{\hat{\beta}} (b) >0 , \dot{\eta}^{\hat{\beta}} (b) = 0.
    \label{64}
\end{eqnarray}
where $\beta = \{r,\theta,\phi\}$. $\eta^{\hat{\beta}} (b)$ represents the separation between two nearby geodesics at $r=b$ in the radial and angular directions. The initial condition in Eq. (\ref{64}) represents a mass released from rest at $r=b$. In the next sections we will discuss the components of radial and the angular deviation vectors in detail.

\subsection{RADIAL COMPONENT}\label{IV A}
\begin{figure*}[htbp]
    \centering
        \includegraphics[width=0.45\linewidth]{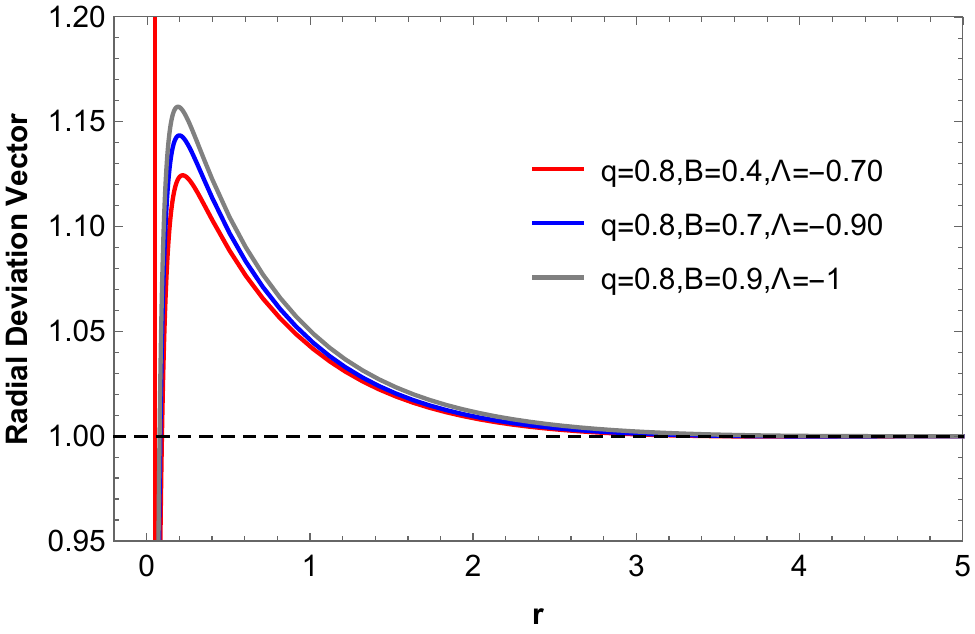}  
        \includegraphics[width=0.45\linewidth]{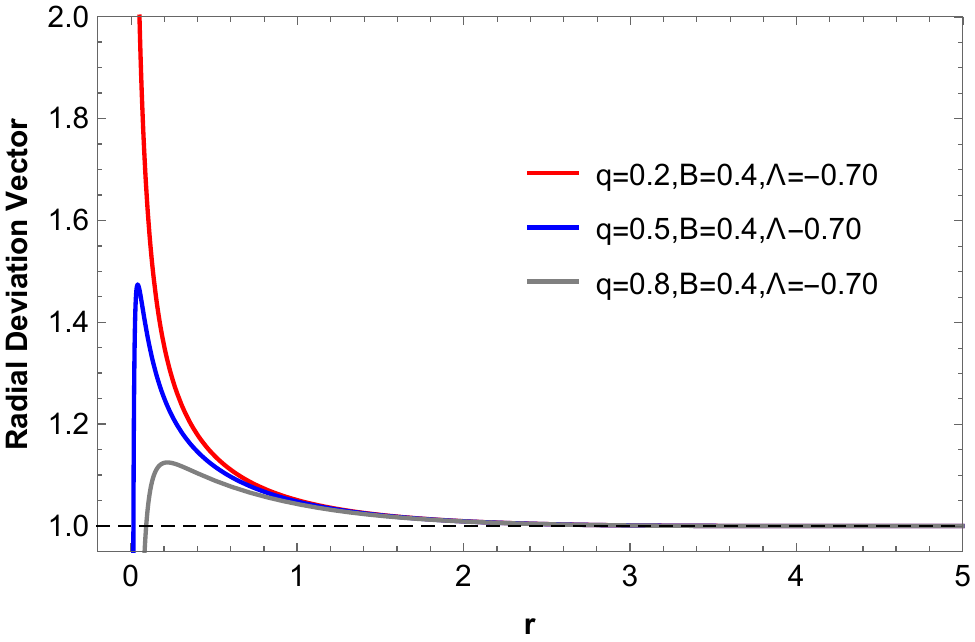}
        \caption{Radial geodesic deviation profile for a radially in-falling particle in AdS black hole spacetime. We use $M=1$ to show the trend for different values of the spacetime parameters $q,B,\Lambda$. We take $b=5M$ as the starting position of the particle and $E=3$ as the energy of the particle at the starting position. }
        \label{fig:12}
\end{figure*}

In Fig.(\ref{fig:12}) we have shown the radial component of the geodesic deviation vector after solving Eq. (\ref{60}) with initial condition. It can be observed that for a fixed value of $q$ and varying values of $B$ and $\Lambda$, the geodesic separation increases as a test particle is released from a point $r=b$. The separation reaches a peak positive value at a particular radial coordinate that lies between the event horizon and the Cauchy horizon. After that, the separation vector starts to decrease rapidly and approaches infinity as $r \rightarrow 0$. It can be observed that increasing the value of $B$ makes the peak value of deviation higher. Fig. (\ref{fig:12}) also shows the deviation trend for different values of $q$, keeping $B$ and $\Lambda$ constant. An opposite behavior is noticed when compared with the case having fixed values of $q$. Here, as the value of $q$ is increased, the peak value of deviation decreases.  
 
\subsection{ANGULAR COMPONENT}\label{IV B}
\begin{figure*}[htbp]
    \centering
        \includegraphics[width=0.45\linewidth]{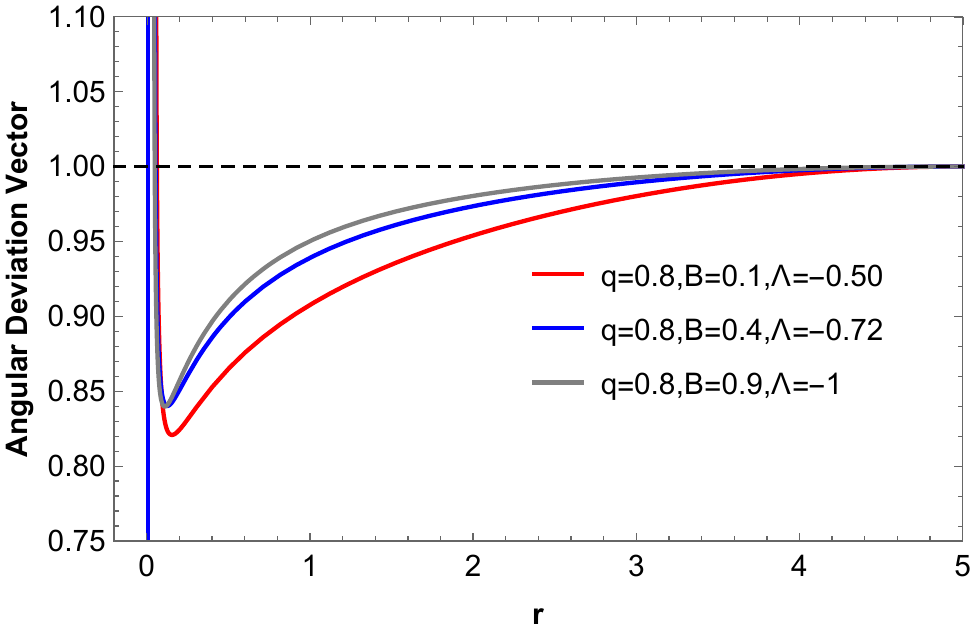}  
         \includegraphics[width=0.45\linewidth]{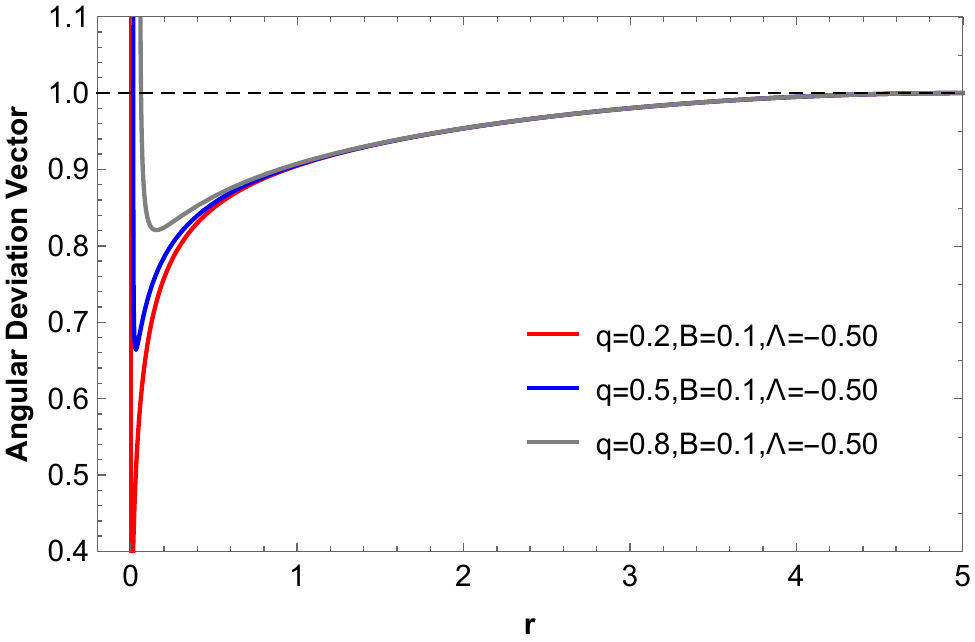}
        \caption{Angular geodesic deviation profile for a radially in-falling particle in AdS black hole spacetime. We use $M=1$ to show the trend for different values of the spacetime parameters $q,B,\Lambda$. We take $b=5M$ as the starting position of the particle and $E=3$ as the energy of the particle at the starting position.}
        \label{fig:13}
    \end{figure*}

Fig. (\ref{fig:13}) shows the angular deviation trend for a radially in-falling particle in AdS black hole using the initial condition in Eq. (\ref{64}). Here, we fix the value of $q=0.8$ and vary the values of $B$ and $\Lambda$. It can be seen that the separation vector decreases monotonically up to a certain peak value as the particle approaches the spacetime singularity. It reverses its trend and starts growing rapidly and then fall to zero as $r \rightarrow 0$. Similar to the case of radial deviation, the angular separation vector also attains the peak value between the Cauchy and the event horizon. It can be noticed that increasing the value of $B$ or equivalently the cosmological constant becomes more negative, the peak value of the separation vector increases as the particle approaches the central spacetime singularity. Fig (\ref{fig:13}) also shows the trend for changing values of $q$, keeping $B$ and $\Lambda$ fixed. Here we can see that increasing the value of $q$ also increases the peak value of deviation attained.

\section{Conclusion}
In this paper, we have considered  a black hole surrounded by Chaplygin dark fluid and studied thermodynamics in the presence of Bekenstien entropy, and examined  the  Joule-Thomson expansion. We studied  the Joule-Thomson coefficient, the inversion temperature and  we also graphed the isenthalpic curves in the $T_i -P_i$ plane.  The Joule-Thomson coefficient is computed, and the inversion curves and the isenthalpic curves are discussed in AdS black holes surrounded by Chaplygin dark fluid.  In the $T -P$ plane, the inversion temperature curves and isenthalpic curves are investigated  with different parameters. In addition, we  obtained the results of behaviors like Van der Waals fluid and the black hole to study their similarities and differences. We have compared our results with recent published papers \cite {J4,J5,J6}.\\\\ 
We have also studied tidal force effects and geodesic deviation for a radially in-falling particle in AdS black hole surrounded by Chaplygin dark fluid. This work extends the pre existing literature on thermodynamics and observational techniques to test various theories of gravity, and aims to distinguish different classes of compact objects theorized in such literature. We have explored the radial and angular tidal forces acting on a particle freely falling towards the AdS black hole. It is noticed that both the tidal force profiles have some finite value as $r \rightarrow \infty$. This behaviour is solely governed by the presence of an effective non-zero cosmological constant. A peak value is attained for both the force profiles after which the tidal force rapidly falls to switch its behaviour i.e radial stretching gets converted into radial compression and vice versa. The peak value is attained between the Cauchy and the event horizon of the black hole. The radial tidal force is seen to become more of stretching nature as the value of $B$ is increased or when the value of $q$ is reduced. Similarly, as the value of $B$ is increased or when $q$ is reduced, the angular tidal force becomes more compressive\\\\
In addition to this, we have also examined the variation of geodesic separation vector with the radial coordinate for two nearby radial geodesics. To numerically solve the geodesic deviation equation we consider the initial condition where the particle is released from rest. We have studied two cases where in the first the value of $q$ is kept as constant and in the second one, the value of $B$ and $\Lambda$ is kept constant. The separation between two adjacent geodesics for both the radial and angular profiles increases for increasing values of $B$ or when the value of the cosmological constant becomes more negative. Similarly, the radial separation vector also increases when the value of $q$ is decreased. The angular separation vector shows opposite trend for changing values of $q$ since it increases for higher values of $q$.
\section*{Acknowledgement}
This project was supported by the natural sciences foundation of
China (Grant No. 11975145). The authors thank the reviewers for
their comments on this paper.
\section*{Declaration of competing interest}
The authors declare that they have no known competing financial interests or personal relationships that could have appeared to
influence the work reported in this paper.
\section*{Data Availability Statement} This manuscript has no associated data, or the data will not be deposited.
(There is no observational data related to this article. The
necessary calculations and graphic discussion can be made available
on request.)


\end{document}